\documentclass[a4paper,11pt]{article}
\usepackage{cite}
\usepackage[all]{xy}
\usepackage{amsmath,amsfonts,amssymb,amsthm,epsfig,amscd,comment,latexsym,psfrag}
\usepackage{CJK}
\usepackage{mathrsfs}
\usepackage{nicefrac,xspace,tikz}
\usepackage{arydshln}
\usepackage{stmaryrd}
\usepackage{graphicx,bm}
\usepackage{amscd}

\numberwithin{equation}{section}
\usetikzlibrary{arrows}

\newcommand{\twoh}
{\begin{tikzpicture}
\draw (0.4,0) rectangle (0.8,0.2);
\draw (0.6,0)--(0.6,0.2);
\end{tikzpicture}}
\newcommand{\twov}
{\begin{tikzpicture}
\draw (0.4,0.2) rectangle (0.6,-0.2);
\draw (0.4,0)--(0.6,0);
\end{tikzpicture}}

\makeatletter

\newcommand{\Rmnum}[1]{\expandafter\@slowromancap\romannumeral #1@}
\makeatother
\usepackage{graphicx,amssymb,amsmath,bm,latexsym}
\allowdisplaybreaks

\begin{document}
\begin{titlepage}
\begin{center}
{\Large\bf CFT approach to constraint operators for ($\beta$-deformed) hermitian one-matrix models}\vskip .2in
{\large Rui Wang$^{a,}$\footnote{wangrui@cumtb.edu.cn},
 Chun-Hong Zhang$^{b,}$\footnote{zhangchunhong@ncwu.edu.cn},
Fu-Hao Zhang$^{c,}$\footnote{zhangfh@cnu.edu.cn},
Wei-Zhong Zhao$^{c,}$\footnote{Corresponding author: zhaowz@cnu.edu.cn}} \vskip .2in
$^a${\em Department of Mathematics, China University of Mining and Technology,
Beijing 100083, China}\\
$^b${\em School of Mathematics and Statistics, North China University of Water Resources and Electric Power,
Zhengzhou 450046, Henan, China}\\
$^c${\em School of Mathematical Sciences, Capital Normal University,
Beijing 100048, China} \\

\begin{abstract}

Since the ($\beta$-deformed) hermitian one-matrix models can be represented as the integrated conformal
field theory (CFT) expectation values, we construct the operators in terms of the generators of the
Heisenberg algebra such that the constraints can be derived by inserting the constructed operators into
the integrated expectation values. We also obtain the second order total derivative operators associating
with the derived constraint operators and analyze their properties. We explore the intrinsic connection
between the derived constraint operators and $W$-representations of some matrix models.
For the Gaussian hermitian one-matrix model in the external field and $\beta$-deformed $N\times N$ complex
matrix model, we investigate the superintegrability and derive the corresponding character expansions
from their $W$-representations. Moreover a conjectured formula for the averages of Jack polynomials in the
literature is proved.

\end{abstract}

\end{center}


\end{titlepage}

\section{Introduction}

The hermitian matrix models have attracted much attention, which correspond to a $D=0$ quantum field theory
where the 'tHooft topological large-$N$ expansion is explicitly solvable. The hermitian matrix models can be
expressed as the eigenvalue models as well. These hermitian eigenvalue models can be reformulated by using the
conformal field theory (CFT) techniques \cite{Marshakov1991}-\cite{Morozov1994}. More precisely, these models
can be defined by the expectation values, which are written completely in terms of CFT quantities.
The CFT formalism can be applied to generate the $1/N$ expansion for any classical solution of the hermitian
one-matrix model. The $1/N$ expansion \cite{tHooft} has wide applications in solving various combinatorial problems
like the enumeration of planar graphs \cite{Brezin1978,tHooft1998}, the dually weighted planar graphs \cite{Kazakov1996},
the branched coverings of a compact Riemann surface \cite{Kostov1996,Kostov1998}, and so on. In terms of a free bosonic
field defined on a Riemann surface determined by the classical background, the $1/N$ expansion can be constructed,
and the spectral correlations are determined directly from the correlation functions of this bosonic field \cite{Kostov}.
The quantum curves arise in matrix models as quantization of spectral curves. As a consequence of the relations
between matrix models and CFT, the singular vector structure of quantum curves follows automatically from the CFT construction.
Moreover, for a given algebraic curve one can associate an infinite family of quantum curves, which are in one-to-one
correspondence with the Virasoro singular vectors \cite{Manabe}. The CFT formalism has been generalized to the construction
of matrix models associated with many other different algebraic structures, such as the deformed matrix
models \cite{Odake2002}-\cite{Nedelin2017} and supersymmetric eigenvalue models \cite{Ciosmak2016,Ciosmak}.
In the case of the super-eigenvalue models, there are the super-quantum curves
which have the structure of the super-Virasoro singular vectors.

The constraints play a fundamental role in understanding the structures of the matrix models. Well known is that the
Virasoro constraints for the hermitian one-matrix model are easily derived from the invariance of partition function
under the proper infinitesimal transformations. In order to give more constraints, Itoyama et al.
\cite{Itoyama} proposed an approach to derive a large class of constraints, namely $W_{1+\infty}$ constraints,
which are associated with the higher order differential operators of the $W_{1+\infty}$ algebra.
The Gaussian hermitian one-matrix model is equivalent to the quadratic Kontsevich-Penner model \cite{AlexandrovJHEP2009,MironovPLB2021}.
It can be expressed as the form of the $W$-representation \cite{Shakirov2009} and possesses the superintegrability
\cite{MironovJHEP082018,2203.03869}, i.e., the average of a properly chosen symmetric function is proportional to ratios of symmetric functions
on a proper locus. The Virasoro constraints for the Gaussian hermitian one-matrix model arise from the Ward identities.
It was shown that the character expansion of this model can be derived by using the Virasoro constraints \cite{Mironov2104,Mironov2105}.
The final result shows that its character expansion is a KP $\tau$-function \cite{Alexandrov2014,AMironov1705}.
Recently, in terms of the Euler and Lassalle operators and the potentials of the $(A_{N-1})$ $B_{N}$-Calogero models,
the Virasoro constraints with higher algebraic structures for the hermitian one-matrix model were constructed \cite{Wangr2019},
where the constraint operators obey the centerless Virasoro algebra and null 3-algebra. There are also the Virasoro constraints
with higher algebraic structures for the fermionic matrix model, and the superintegrability can be easily derived from such
Virasoro constraints \cite{LY}.

In this paper, we reinvestigate the ($\beta$-deformed) hermitian one-matrix models given by the integrated CFT expectation values.
One purpose of this paper is to construct the new constraints and explore the intrinsic connection between the constraint operators
and $W$-representations of some matrix models. We also analyze the character expansions of some matrix models from $W$-representations.
Another purpose is to derive the total derivative operators with respect to the integration variables associating with the constraint
operators and present their remarkable properties.

This paper is organized as follows. In section 2, we recall the hermitian one-matrix model given by an integrated
CFT expectation value. Then we construct the operators in terms of the generators of the Heisenberg algebra.
By means of these operators, we derive the constraints for the hermitian one-matrix model which contain
the $A$ and $B$-type Lassalle constraints. We explore the intrinsic connection between the derived constraint operators
and $W$-representations of some matrix models. We analyze the character expansions of matrix modes based on $W$-representations.
In addition, the generalized Virasoro constraints with higher algebraic structures are also presented.
In section 3, we derive the similar constraints for the $\beta$-deformed matrix model and present the derived total derivative
operators associating with the constraint operators. We analyze the derived total derivative operators and extend them to more general
operators. The remarkable properties of these operators are presented. For the derived constraint operators, we show that two
of them are associated with the $W$-representations of $\beta$-deformed Gaussian hermitian and $N\times N$ complex matrix models.
We derive the superintegrability of $\beta$-deformed $N\times N$ complex matrix model from the $W$-representation.
We end this paper with the conclusions in section 4.

\section{CFT approach to constraint operators in the hermitian one-matrix model}
\subsection{CFT approach to constraint operators in the hermitian one-matrix model}

Let us introduce the local bosonic field
\begin{equation}
\phi(z)=\phi_{+}(z)+\phi_{-}(z),
\end{equation}
where $\phi_{\pm}(z)$ are given by
\begin{eqnarray}\label{phi}
&&\phi_{+}(z)=a_0{\rm{log}}z-\sum_{m=1}^{\infty}\frac{a_m}{m}z^{-m},\nonumber\\
&&\phi_{-}(z)=q+\sum_{m=1}^{\infty}\frac{a_{-m}}{m}z^{m},
\end{eqnarray}
in which the operators $a_m$ yield the Heisenberg algebra
\begin{equation}\label{Halg}
[a_m, a_n]=\frac{1}{2}m\delta_{m+n,0},
\end{equation}
and the operator $q$ satisfies
\begin{equation}
[a_m, q]=\frac{1}{2}\delta_{m,0}.
\end{equation}
The commutator of $\phi_{+}(z)$ and  $\phi_{-}(z)$ is
\begin{equation}\label{commutatorphi}
[\phi_{+}(z_1), \phi_{-}(z_2)]=\frac{1}{2}{\rm log}(z_1-z_2),\quad
|z_1|>|z_2|.
\end{equation}

In terms of the bosonic field $\phi(z)$, the energy momentum tensor is given by
\begin{equation}
T(z)=:\partial \phi(z)\partial \phi(z):=\sum_{m\in\mathbb{Z}}L_mz^{-m-2}.
\end{equation}
The modes $L_{m}$ are explicitly written as
\begin{equation}
L_m=\sum_{n\in \mathbb{Z}}:a_{m-n}a_n:,
\end{equation}
which yield the Virasoro algebra
\begin{equation}
[L_m,L_n]=(m-n)L_{m+n}+\frac{1}{12}(m^3-m)\delta_{m+n,0},\quad m,n\in \mathbb{Z}.
\end{equation}
The actions of the generators $a_m$ on the vacuum state $|0\rangle$ are defined by
\begin{equation}\label{state}
a_m|0\rangle=0,\quad m\geq 0.
\end{equation}
Then we have
\begin{equation}\label{Lcon}
L_{m}|0\rangle=0,\quad m\geq -1.
\end{equation}

We work with the vertex operators
$V^{-1}(z)=e^{-2\phi_{-}(z)}e^{-2\phi_{+}(z)}$,
they are the primary fields with respect to the Virasoro algebra, i.e.,
\begin{equation}\label{LE}
[L_m, V^{-1}(z)]=((m+1)z^m+z^{m+1}\frac{\partial}{\partial z})V^{-1}(z).
\end{equation}

Note that
\begin{equation}\label{vander}
\prod_{i=1}^{N}V^{-1}(z_i)|0\rangle=\Delta(z)^{2}
e^{-2\sum_{i=1}^{N}\phi_{-}(z_i)}|0\rangle,
\end{equation}
and
\begin{equation}\label{actionv}
\langle U_{N,t}|e^{-2\sum_{i=1}^{N}\phi_{-}(z_i)}=e^{\sum_{i=1}^{N}U(z_i)}
\langle U_{0,t}|,
\end{equation}
where $\Delta(z)=\prod\limits_{1\leq i<j\leq N}(z_i-z_j)$ is the Vandermonde determinant,
$U(z)=\sum_{m=0}^{\infty}t_mz^m$,
the coherent bra state $\langle U_{N,t}|$ is
$\langle U_{N,t}|=\langle 0|e^{2Nq}e^{-\sum_{m=0}^{\infty}t_ma_m}$.

The hermitian one-matrix model can be represented as an integrated CFT expectation value \cite{Marshakov1991}
\begin{equation}\label{PF}
Z=\prod_{i=1}^N\int_{-\infty}^{+\infty}  dz_i \langle U_{N,t}|\prod_{i=1}^{N}V^{-1}(z_i)|0\rangle
=\prod_{i=1}^N\int_{-\infty}^{+\infty}  dz_i \Delta(z)^2 e^{\sum_{i=1}^{N}U(z_i)}.
\end{equation}
There are the Virasoro constraints
$\hat L_m Z=0,\ \ m\geq -1$,
with
\begin{equation}\label{virconop}
\hat L_m=\sum_{k=1}^{\infty}kt_k\frac{\partial}{\partial t_{k+m}}+
\sum_{k=0}^{m}\frac{\partial}{\partial t_{k}}
\frac{\partial}{\partial t_{m-k}}.
\end{equation}

Let us now construct the operator
\begin{equation}
\tilde W(z)=-2\partial \phi_{-}(z)T(z)=\sum_{n\in\mathbb{Z}}\tilde W_nz^{-n-3},
\end{equation}
then the modes $\tilde W_n$ are given by
\begin{equation}\label{modewn}
\tilde W_n=-2\sum_{m=1}^{\infty}a_{-m}L_{m+n}.
\end{equation}
Note that
\begin{equation}\label{WnL0}
[\tilde W_{n},L_0]=n\tilde W_{n}.
\end{equation}
There are the similar expressions with (\ref{WnL0})
\begin{equation}\label{LnL0}
[L_n,L_0]=nL_{n},\quad n\in \mathbb{Z},
\end{equation}
which play an important role in the construction of the well known highest weight representations
of the Virasoro algebra.

In terms of the operators $L_0$ and $\tilde W_{n}$ with any given $n$, we may construct the operators
${\tilde L}_m=\frac{1}{n}\tilde W_{n}^m L_0$, $m\in \mathbb{N}$.
They yield the centerless Virasoro algebra  and null 3-algebra
\begin{equation}\label{nullalg}
[{\tilde L}_k, {\tilde L}_l, {\tilde L}_m]={\tilde L}_k[{\tilde L}_l, {\tilde L}_m]
+{\tilde L}_l[{\tilde L}_m, {\tilde L}_k]
+{\tilde L}_m[{\tilde L}_k, {\tilde L}_l]
=0.
\end{equation}

When $n\geq -2$ in (\ref{modewn}),  we write the operators $\tilde W_n$ as
\begin{equation}\label{Mna}
\tilde W_{n}=-4\sum_{k,l=1}^{\infty}a_{-k}a_{-l}a_{k+l+n}
-2\sum_{k=1}^{\infty}\sum_{l=0}^{k+n}a_{-k}a_la_{k-l+n}.
\end{equation}
Inserting (\ref{Mna}) into the expectation value in (\ref{PF}), we obtain
\begin{eqnarray}\label{Mncon}
\prod_{i=1}^N\int_{-\infty}^{+\infty} dz_i \langle U_{N,t}|\tilde W_{n}\prod_{i=1}^{N}V^{-1}(z_i)|0\rangle
=\prod_{i=1}^N\int_{-\infty}^{+\infty} dz_i {\bar W}_{n}\langle U_{N,t}|\prod_{i=1}^{N}V^{-1}(z_i)|0\rangle=0,
\end{eqnarray}
where the operators $\bar W_{n}$ are given by
\begin{equation}\label{Mnz}
\bar W_{n}=\sum_{i=1}^{N}\frac{\partial^2}{\partial z^2_i}z^{n+2}_i
-2\sum_{i=1}^{N}\sum_{j\neq i}\frac{\partial}{\partial z_i}\frac{z^{n+2}_i}{z_i-z_j}
-(n+2)\sum_{i=1}^{N}\frac{\partial}{\partial z_i}z^{n+1}_i, \ \ n\geq -2.
\end{equation}

Then we may derive the constraints from (\ref{Mncon})
\begin{equation}
\hat W_nZ=0,\ \ n\geq -2,
\end{equation}
where
\begin{equation}\label{Mnt}
\hat W_{n}=\sum_{k,l=1}^{\infty}klt_kt_l\frac{\partial}{\partial t_{k+l+n}}
+\sum_{k=1}^{\infty}\sum_{l=0}^{k+n}kt_k\frac{\partial}{\partial t_l}\frac{\partial}{\partial t_{k-l+n}}.
\end{equation}
Taking $n=-2$ and $-1$ in (\ref{Mnt}), respectively, we have the $A$ and $B$-type Lassalle constraints \cite{Wangr2019}
\begin{equation}\label{M2t}
\hat W_{-2}=\sum_{k,l=1}^{\infty}klt_kt_l\frac{\partial}{\partial t_{k+l-2}}
+\sum_{k=2}^{\infty}\sum_{l=0}^{k-2}kt_k\frac{\partial}{\partial t_l}\frac{\partial}{\partial t_{k-l-2}},
\end{equation}
\begin{equation}\label{M1t}
\hat W_{-1}=\sum_{k,l=1}^{\infty}klt_kt_l\frac{\partial}{\partial t_{k+l-1}}
+\sum_{k=1}^{\infty}\sum_{l=0}^{k-1}kt_k\frac{\partial}{\partial t_l}\frac{\partial}{\partial t_{k-l-1}}.
\end{equation}

In terms of the constraint operators $\hat L_0$ and $\hat W_{n}$ with any given $n$,
we may construct the Virasoro constraints
\begin{equation}\label{V3con}
{\hat {\mathbb{L}}}_mZ=0,
\end{equation}
with ${\hat {\mathbb{L}}}_m=\frac{1}{n}\hat W_{n}^m\hat L_0,$
$m\in \mathbb{N}$.
The operators ${\hat {\mathbb{L}}}_m$ constitute the centerless Virasoro algebra and null 3-algebra
which are isomorphic to the algebras generated by the operators ${\tilde L}_m$.

Let us turn to consider the commutator of (\ref{Mna}) and $L_m$ with $m\geq -1$,
\begin{equation}\label{WnLm}
[\tilde W_{n},L_m]=
(n-2m)\tilde W_{m+n}
+2\sum_{k=0}^{m-1}(m-k)L^{m+n}_k,
\end{equation}
where $L^{m+n}_k=a_kL_{m+n-k}$.

The operators $L^{m+n}_{k}$ can be extended to more general cases
\begin{equation}\label{genvir}
L^m_{k_1,\cdots,k_r}=a_{k_1}\cdots a_{k_r}L_m, \ \ m\geq -1, k_1,\cdots,k_r\geq 0,
\end{equation}
such that they yield the closed algebra
\begin{eqnarray}\label{genviralg}
[L^m_{k_1,\cdots,k_r},L^n_{l_1,\cdots,l_s}]
&=&(m-n)L^{m+n}_{k_1,\cdots,k_r,l_1,\cdots,l_s}+\sum_{i=1}^rk_i
L^m_{k_1,\cdots,k_{i-1},k_i+n,k_{i+1},\cdots,k_r,l_1,\cdots,l_s}\nonumber\\
&&-\sum_{j=1}^sl_jL^n_{k_1,\cdots,k_r,l_1,\cdots,l_{j-1},l_j+m,l_{j+1},\cdots,l_s},
\end{eqnarray}
and null 3-algebra
\begin{equation}\label{genvirnullalg}
[L^m_{k_1,\cdots,k_r},L^m_{l_1,\cdots,l_s},L^m_{n_1,\cdots,n_t}]=0.
\end{equation}
When particularized to the $k_i=0$, $l_j=0$, $i=1,\cdots,r$ and $j=1,\cdots,s$ case in (\ref{genviralg}),
it reduces to the centerless Virasoro algebra.

Similarly to (\ref{Mncon}), we have
\begin{eqnarray}\label{genvirconid}
&&\prod_{i=1}^N\int_{-\infty}^{+\infty} dz_i \langle U_{N,t}|L^m_{k_1,\cdots,k_r}\prod_{i=1}^{N}V^{-1}(z_i)|0\rangle\nonumber\\
&=&\prod_{i=1}^N\int_{-\infty}^{+\infty} dz_i \bar L^m_{k_1,\cdots,k_r}\langle U_{N,t}|\prod_{i=1}^{N}V^{-1}(z_i)|0\rangle=0,
\end{eqnarray}
where the operators $\bar L^m_{k_1,\cdots,k_r}$ are given by
$\bar L^m_{k_1,\cdots,k_r}=(-1)^r\sum_{i,j_1,\cdots, j_r=1}^{N}\frac{\partial}{\partial z_i}z_i^{m+1}z_{j_1}^{k_1}\cdots z_{j_r}^{k_r}$.

Then we may derive the generalized Virasoro constraints from (\ref{genvirconid})
\begin{equation}\label{genvircon}
\hat L^m_{k_1,\cdots,k_r}Z=0,\ \ m\geq -1,
\end{equation}
where
\begin{eqnarray}\label{genvirconops}
\hat L^m_{k_1,\cdots,k_r}&=&\sum_{k=1}^{\infty}kt_k\frac{\partial}{\partial t_{k+m}}\frac{\partial}{\partial t_{k_1}}\cdots
\frac{\partial}{\partial t_{k_r}}+\sum_{k=0}^{m}\frac{\partial}{\partial t_{k}}
\frac{\partial}{\partial t_{m-k}}\frac{\partial}{\partial t_{k_1}}\cdots \frac{\partial}{\partial t_{k_r}}\nonumber\\
&&+\sum_{i=1}^{r}k_i\frac{\partial}{\partial t_{k_1}}\cdots  \frac{\partial}{\partial t_{k_{i-1}}}\frac{\partial}{\partial t_{m+k_i}}
\frac{\partial}{\partial t_{k_{i+1}}}\cdots \frac{\partial}{\partial t_{k_r}},
\end{eqnarray}
 and the constraint operators (\ref{genvirconops}) yield  the generalized centerless Virasoro algebra (\ref{genviralg}) and  the
null 3-algebra (\ref{genvirnullalg}). The constraints (\ref{genvircon}) were also derived in Ref. \cite{DXM} from the  multi-loop
equations, while the null 3-algebra was not mentioned there.

\subsection{$W$-representations of the matrix models}
$W$-representations of the matrix models may realize the partition functions by acting on elementary functions with exponents
of the given $W$-operators. They indeed give the dual expressions for the partition functions through differentiation rather
than integration. To show the connection between $\hat W_{n}$ (\ref{Mnt}) and $W$-representations of some matrix models,
let us take the rescaling variables $p_k=kt_k$ $(k>0)$, and substitute $\frac{\partial}{\partial t_0}$ by $N$ in (\ref{Mnt}),
then we have
\begin{eqnarray}\label{hMnp}
{\hat W}_{n}&=&\sum_{k,l=1}^{\infty}\big((k+l+n)p_{k}p_l
\frac{\partial}{\partial p_{k+l+n}}+klp_{k+l-n}\frac{\partial}{\partial p_k}\frac{\partial}{\partial p_l}\big)\nonumber\\
&&+2N\sum_{k=1}^{\infty}(k+n)p_{k}
\frac{\partial}{\partial p_{k+n}}+N(p_1^2+Np_2)\delta_{n,-2}+N^2p_1\delta_{n,-1}, \quad n\geq -2.
\end{eqnarray}
Here we have set $p_k=\frac{\partial}{\partial p_{k}}=0$ for $k\leq 0$.

We see that the operators $\hat W_{-2}$, $\hat W_{-1}$, $\hat W_{0}$ and
$\hat W_{2}$ appear in the $W$-representations of the following matrix models:

(i) Gaussian hermitian one-matrix model \cite{Shakirov2009,AMironov1705}
\begin{eqnarray}\label{GPF}
Z_G\{p\}&=&2^{-\frac{N}{2}}\pi^{-\frac{N^2}{2}}\int_{N\times N} dM e^{-\frac{1}{2}{\rm Tr}M^{2}+\sum_{k=1}^{\infty}\frac{p_k}{k}{\rm Tr}M^k}\nonumber\\
&=&e^{{\hat W}_{-2}/2}\cdot 1=\sum_{\lambda} S_{\lambda}\{p_k=\delta_{k,2}\}\frac{D_{\lambda}(N)}{d_\lambda}S_{\lambda}\{p\},
\end{eqnarray}
where $\lambda=(\lambda_1, \lambda_2,\cdots, \lambda_{l(\lambda)})$, $\lambda_i\in \mathbb{N}_{+}$, $\lambda_i\geq \lambda_{i+1}$
is the partition, $l(\lambda)$ is the length of the partition $\lambda$,
$S_{\lambda}\{p\}$ are the Schur functions, $D_{\lambda}(N)=S_{\lambda}\{p_k=N\}$ and $d_{\lambda}=S_{\lambda}\{p_k=\delta_{k,1}\}$~
are respectively the dimension of representation~$\lambda$~
for the linear group~$GL(N)$ and symmetric group $S_{|\lambda|}$~divided by~$|\lambda|!$ \cite{Fulton}.

(ii) $N\times N$ complex matrix model \cite{AlexandrovJHEP2009,AMironov1705}
\begin{eqnarray}\label{CPF}
Z_{N\times N}\{p\}&=&\frac{\int_{N\times N} d^2M e^{-{\rm Tr}MM^{\dagger}+\sum_{k=1}^{\infty}\frac{p_k}{k}{\rm Tr}(MM^{\dagger})^k}}
{\int_{N\times N} d^2M e^{-{\rm Tr}MM^{\dagger}}}\nonumber\\
&=&e^{{\hat W}_{-1}}\cdot 1=\sum_{\lambda}\frac{D_{\lambda}(N)^2}{d_\lambda}S_{\lambda}\{p\}.
\end{eqnarray}

(iii) Hurwitz-Kontsevich matrix model \cite{Goulden97,KHurwitz,Shakirov2009}
\begin{eqnarray}\label{HKPF}
Z_{HK}\{p\}=e^{t{\hat W}_{0}/2}\cdot e^{p_1/e^{tN}}
=\sum_{\lambda}e^{tc_{\lambda}}S_\lambda\{p_k=e^{-tN}\} S_{\lambda}\{p\},
\end{eqnarray}
where $t$ is a deformation
parameter and  $c_{\lambda}=\frac{1}{2}\sum_{i=1}^{l(\lambda)}\lambda_i(\lambda_i-2i+1)$.

(iv) Hermitian one-matrix model in the external field \cite{Shakirov2009}
\begin{eqnarray}\label{HEPF}
Z\{g|p\}&=&\int dM_1 e^{-{\rm Tr}\frac{M_1^{2}}{2}+\sum_{k=1}^{\infty}\frac{g_k}{k}{\rm Tr}(M_1+M_2)^k}\nonumber\\
&=&e^{\mathscr{W}_2/2}
\cdot e^{\sum_{k=1}^{\infty}\frac{1}{k}g_kp_k},
\end{eqnarray}
where $\mathscr{W}_2={\hat W}_{2}+2N^2\frac{\partial}{\partial p_{2}}+N\frac{\partial^2}{\partial p^2_{1}}$, $p_k={\rm Tr}M_2^k$.

In what follows, we consider the actions of the operators (\ref{hMnp}) on the Schur functions $S_{\lambda}$.
Firstly, we notice that $\hat W_0=2\mathscr{W}_0+2N\mathscr{D}$,
where $\mathscr{W}_0$  and $\mathscr{D}$ are the cut-and-join operators \cite{cutjoin}
\begin{eqnarray}
\mathscr{W}_0&=&\frac{1}{2}\sum_{k,l=1}^{\infty}\big((k+l)p_{k}p_l
\frac{\partial}{\partial p_{k+l}}+klp_{k+l}\frac{\partial}{\partial p_k}\frac{\partial}{\partial p_l}\big),\nonumber\\
\mathscr{D}&=&\sum_{k=1}^{\infty}kp_{k}\frac{\partial}{\partial p_k}.
\end{eqnarray}

Due to the actions
\begin{eqnarray}
\mathscr{W}_0S_{\lambda}&=&c_{\lambda}S_{\lambda},
\label{W0action}\\
\mathscr{D}S_{\lambda}&=&|\lambda|S_{\lambda},\label{Daction}
\end{eqnarray}
where $|\lambda|=\sum_{i=1}^{l(\lambda)}\lambda_i$, we have
\begin{equation}
\hat W_0S_{\lambda}=\sum_{i=1}^{l(\lambda)}\lambda_i(\lambda_i-2i+1+2N)S_{\lambda}.
\end{equation}

In order to obtain the actions $\hat W_{ n}S_{\lambda}$ $(n\neq 0)$,
we rewrite the operators $\hat W_{n}$ as
\begin{eqnarray}\label{chMnp}
{\hat W}_{n}=
\left\{
  \begin{array}{l}
    \frac{1}{n}[\mathscr{W}_0,[\mathscr{W}_0, \frac{\partial}{\partial p_n}]]-2N[\mathscr{W}_0,
    \frac{\partial}{\partial p_n}]-\frac{1}{12}n(n^2-1)\frac{\partial}{\partial p_n}
-N\sum_{k=1}^{n-1}k(n-k)\frac{\partial}{\partial p_{k}}
\frac{\partial}{\partial p_{n-k}}
\\-\frac{1}{2n}\sum_{k,l=1}^{n-2}kl(k+l)(n-k-l)
\frac{\partial}{\partial p_k}\frac{\partial}{\partial p_l}\frac{\partial}{\partial p_{n-k-l}}, \quad n>0; \\
\\
    \frac{1}{n^2}[\mathscr{W}_0,[\mathscr{W}_0, p_{-n}]]-\frac{2N}{n}[\mathscr{W}_0, p_{-n}]
+(N^2-\frac{1}{12}(n^2-1))p_{-n},  \quad n=-1,-2.
  \end{array}
\right.
\end{eqnarray}
Using the action (\ref{W0action}) and \cite{Macdonaldbook}
\begin{eqnarray}
p_nS_{\lambda}&=&\sum_{k=1}^{n}\sum_{\mu}
(-1)^{k}a^{\mu}_{\lambda,(k,1^{n-k})}S_{\mu},\label{multiply}\\
n\frac{\partial}{\partial p_n}S_{\lambda}&=&\sum_{k=1}^{n}\sum_{\mu}
(-1)^{k}a^{\lambda}_{\mu,(k,1^{n-k})}S_{\mu},\label{partial}
\end{eqnarray}
where $a^{\lambda}_{\mu,\nu}$ are the Littlewood-Richardson coefficients defined by
$S_{\mu}S_{\nu}=\sum_{\lambda}a^{\lambda}_{\mu,\nu}S_{\lambda}$,
we can in principle derive the actions $\hat W_{ n}S_{\lambda}$ $(n\neq 0)$.

Let us list several action results
\begin{subequations}\label{actionmn}
\begin{eqnarray}
\hat W_{-1}S_{\lambda}&=&\sum_{\lambda+\square}(j_\square-i_\square+N)^2 S_{\lambda+\square},\\
\hat W_{-2}S_{\lambda}&=&\sum_{\lambda+\twoh}(j_\square-i_\square+N)(j_\square-i_\square+1+N)
S_{\lambda+\twoh}
\nonumber\\
&&-\sum_{\lambda+\twov}(j_\square-i_\square+N)(j_\square-i_\square-1+N)
S_{\lambda+\twov},\label{actionmf2}\\
\hat W_1S_{\lambda}&=&\sum_{\lambda-\square}(j_\square-i_\square)(j_\square-i_\square+2N) S_{\lambda-\square},\\
\hat W_2S_{\lambda}&=&\sum_{\lambda-\twoh}(j_\square-i_\square)(j_\square-i_\square+1+2N)
S_{\lambda-\twoh}\nonumber\\
&&-\sum_{\lambda-\twov}(j_\square-i_\square-1)(j_\square-i_\square+2N)
S_{\lambda-\twov}-N\sum_{\square\neq \square'}S_{\lambda-\square-\square'},\label{actionmz2}
\end{eqnarray}
\end{subequations}
where  $\lambda\pm \square$, $\lambda\pm\twoh$ and $\lambda\pm\twov$ are respectively the Young diagrams obtained
by adding (removing) one square, two horizontally connected squares and two vertically connected squares to (from)
$\lambda$, $\lambda-\square-\square'$ is the Young diagram obtained by removing two different squares
from $\lambda$, $(i_\square, j_\square)$ are the coordinates of the squares added to or removed from  $\lambda$.

We list out the character expansions of the matrix models (\ref{GPF})-(\ref{HKPF}) in the literature.
The character expansion of the Gaussian hermitian one-matrix model can be derived recursively from
from a single $w$-constraint \cite{Mironov2105}
\begin{eqnarray}\label{CEGHM}
(\hat W_{-2}-\mathscr{D})Z_{G}\{p\}=0,
\end{eqnarray}
where the partition function is rewritten as
\begin{equation}
Z_{G}\{p\}=\sum_{\lambda}{r}_{\lambda}S_{\lambda}\{p\}.
\end{equation}
More precisely, there are the recursive relations from (\ref{CEGHM})
\begin{eqnarray}\label{RRGHM}
|\lambda|r_{\lambda}&=&\sum_{\lambda-\twoh}(j_\square-i_\square+N)(j_\square-i_\square+1+N)
r_{\lambda-\twoh}
\nonumber\\
&&-\sum_{\lambda-\twov}(j_\square-i_\square+N)(j_\square-i_\square-1+N)
r_{\lambda-\twov}.
\end{eqnarray}
Thus the coefficients  $r_{\lambda}$ can be solved recursively from (\ref{RRGHM}).

We have presented the Virasoro constraints (\ref{V3con}) with higher algebraic structures for
the hermitian one-matrix model. The similar Virasoro constraints for the Gaussian hermitan one-matrix
model are \cite{Kangbei}
\begin{equation}\label{VGcon}
{\hat {\mathbb{L}}}^G_mZ_G\{p\}=0,
\end{equation}
where the constraint operators ${\hat {\mathbb{L}}}^G_m=\frac{1}{2}\hat W_{-2}^{m}(\hat W_{-2}-\mathscr{D}),$
$m\in \mathbb{N}$ obey the centerless Virasoro algebra and null 3-algebra (\ref{nullalg}).

Similar to the case of the fermionic matrix model \cite{LY},
we may obtain the recursions from (\ref{VGcon})
\begin{eqnarray}\label{recursive}
r_{\lambda}=\frac{2^{m+1}(m+1)!}{\prod_{j=0}^{m}(|\lambda|-2j)}
\sum_{\mu,|\lambda/\mu|=2m+2}r_{\mu}S_{\lambda/\mu}\{p_k=\delta_{k,2}\}
\prod_{(i_\square, j_\square)\in \lambda/\mu}
(j_{\square}-i_{\square}+N),
\end{eqnarray}
where $S_{\lambda/\mu}$ are the skew Schur functions.

Then by using the hook formulas
$\frac{D_{\lambda}(N)}{d_\lambda}=\prod_{(i,j)\in \lambda}
(j-i+N)$,
the desired $r_{\lambda}$ can be easily obtained from (\ref{recursive}) with initial value $r_{\varnothing}=1$.

We note that for the case of the Gaussian hermitian one-matrix model in the external field (\ref{HEPF}),
to our best knowledge, its character expansion has not been reported so far in the existing literature.
Let us now derive its character expansion from the viewpoint of $W$-representation.

Using the actions (\ref{partial}) and (\ref{actionmz2}), we have
\begin{eqnarray}\label{actionw2}
{\mathscr W}_2S_{\lambda}&=&\sum_{\lambda-\twoh}(j_\square-i_\square+N)(j_\square-i_\square+1+N)
S_{\lambda-\twoh}\nonumber\\
&&-\sum_{\lambda-\twov}(j_\square-i_\square+N)(j_\square-i_\square-1+N)
S_{\lambda-\twov}.
\end{eqnarray}
Comparing with the action (\ref{actionmf2}), we see that the action (\ref{actionw2}) is removing squares
from $\lambda$ rather than adding squares to $\lambda$.
It is not difficult to obtain
\begin{eqnarray}\label{actionw2m}
{\mathscr W}^m_2S_{\lambda}&=&\sum_{\mu}(-1)^{m-a}n(\mu)\prod_{k=1}^{a}(j_{\square_k}-i_{\square_k}+N)(j_{\square_k}
-i_{\square_k}+1+N)\nonumber\\
&&\prod_{l=1}^{m-a}(j_{\square_l}-i_{\square_l}+N)(j_{\square_l}-i_{\square_l}-1+N)
S_{\mu},\nonumber\\
&=&\sum_{\mu}(-1)^{m-a}n(\mu)\frac{d_{\mu}D_{\lambda}(N)}{d_\lambda D_\mu(N)}S_{\mu},
\end{eqnarray}
where the sum is over all the possible Young diagrams $\mu$ obtained by removing $a$ blocks of ~\twoh~
and $m-a$ blocks of ~\twov~  from $\lambda$, $n(\mu)$ is the number of the different ways from $\lambda$ to $\mu$.

In order to calculate $(-1)^{m-a}n(\mu)$, we observe that there is the similar action
\begin{eqnarray}\label{actionp2}
(2\frac{\partial}{\partial p_2})^mS_{\lambda}=\sum_{\mu}(-1)^{m-a}n(\mu)
S_{\mu}.
\end{eqnarray}
Due to the orthogonality of the Schur functions $\langle S_\lambda, S_{\mu} \rangle=\delta_{\lambda,\mu}$,
we have
\begin{eqnarray}\label{nmu1}
(-1)^{m-a}n(\mu)=\langle(2\frac{\partial}{\partial p_2})^mS_{\lambda}, S_{\mu}\rangle.
\end{eqnarray}
Here the scalar product $\langle \ .\ , \ .\  \rangle$ is given by
\begin{equation}
\langle p_{\lambda}, p_{\mu}\rangle=\delta_{\lambda,\mu}z_{\lambda},
\end{equation}
where $p_{\lambda}=p_{\lambda_1}p_{\lambda_2}\cdots p_{\lambda_{l(\lambda)}}$, $z_{\lambda}=\prod_{i\geq 1}i^{m_i}m_i!$
and $m_i=m_i(\lambda)$ is the number of parts of $\lambda$ equal to $i$.
By the duality $\langle k\frac{\partial}{\partial p_k}S_\lambda, S_{\mu}\rangle=\langle S_\lambda, p_k S_{\mu}\rangle$,
(\ref{nmu1}) can be expressed as
\begin{eqnarray}\label{nmu}
(-1)^{m-a}n(\mu)
=\langle S_{\lambda}, p_2^mS_{\mu}\rangle
=2^mm!S_{\lambda/\mu}\{p_k=\delta_{k,2}\},
\end{eqnarray}
where $S_{\lambda/\mu}$ are the skew Schur functions.

Substituting (\ref{nmu}) into (\ref{actionw2m})
and using the Cauchy formula $e^{\sum_{k=1}^{\infty}\frac{1}{k}p_k\bar p_k}=\sum_{\lambda}S_{\lambda}\{p\}S_{\lambda}\{\bar p\}$,
we finally obtain the character expansion of (\ref{HEPF})
\begin{eqnarray}\label{cexpansion}
Z\{g|p\}=\sum_{\lambda,\mu}S_{\lambda/\mu}\{p_k=\delta_{k,2}\}
\frac{d_{\mu}D_{\lambda}(N)}{d_\lambda D_\mu(N)}S_{\mu}\{p\}S_{\lambda}\{g\}.
\end{eqnarray}

Note that for the Gaussian hermitian one-matrix model, $N\times N$ complex and Hurwitz-Kontsevich matrix models,
the corresponding character expansions can be derived from their $W$-representations as well.

\section{CFT approach to constraint operators in the $\beta$-deformed hermitian matrix model}
\subsection{CFT approach to constraint operators in the $\beta$-deformed hermitian matrix model}
The $\beta$-deformed hermitian matrix model can also be represented as an integrated CFT expectation value \cite{Ciosmak}
\begin{equation}\label{DPF}
Z_{\beta}=\prod_{i=1}^N\int_{-\infty}^{+\infty} dz_i\langle U_{N\sqrt{\beta},t}|\prod_{i=1}^{N}V^{-\sqrt{\beta}}(z_i)|0\rangle
=\prod_{i=1}^N\int_{-\infty}^{+\infty} dz_i \Delta(z)^{2\beta} e^{\sum_{i=1}^{N}U(z_i)},
\end{equation}
where $V^{-\sqrt{\beta}}(z)=e^{-2\sqrt{\beta}\phi_{-}(z)}e^{-2\sqrt{\beta}\phi_{+}(z)}$,
$\langle U_{N\sqrt{\beta},t}|=\langle 0|e^{2N\sqrt{\beta}q}e^{-\frac{1}{\sqrt{\beta}}\sum_{m=0}^{\infty}t_ma_m}$.

In this case, the deformed energy momentum tensor is
\begin{equation}
\mathcal{T}(z)=:\partial \phi(z)\partial \phi(z):+Q\partial^2 \phi(z)=\sum_{m\in\mathbb{Z}}\mathcal{L}_mz^{-m-2},
\end{equation}
where $Q=\sqrt{\beta}^{-1}-\sqrt{\beta}$.
The modes $\mathcal{L}_{m}$ are explicitly written as
\begin{equation}\label{deformvirmod}
\mathcal{L}_m=\sum_{n\in \mathbb{Z}}:a_{m-n}a_n:-(m+1)Qa_m.
\end{equation}
Inserting (\ref{deformvirmod}) with $m\geq -1$ into the expectation value in (\ref{DPF}), we obtain the Virasoro constraints
$$\hat {\mathcal{L}}_{m}Z_{\beta}=0,\ \ \ m\geq -1,$$
where
\begin{eqnarray}
\hat {\mathcal{L}}_{m}=\sum_{k=1}^{\infty}kt_k\frac{\partial}{\partial t_{k+m}}+
\beta\sum_{k=0}^{m}\frac{\partial}{\partial t_{k}}
\frac{\partial}{\partial t_{m-k}}+(1-\beta)(m+1)\frac{\partial}{\partial t_{m}}.
\end{eqnarray}

Let us construct the operators
\begin{equation}
\mathcal{\tilde W}(z)=-2\partial \phi_{-}(z)\mathcal{T}(z)=\sum_{n\in\mathbb{Z}}\mathcal{\tilde W}_nz^{-n-3},
\end{equation}
where the modes $\mathcal{\tilde W}_n$ are given by
\begin{equation}\label{dmodewn}
\mathcal{\tilde W}_n=-2\sum_{m=1}^{\infty}a_{-m}\mathcal{L}_{m+n}.
\end{equation}

When $n\geq -2$ in (\ref{dmodewn}), we may write $\mathcal{\tilde W}_n$ as
\begin{equation}\label{DMna}
\mathcal{\tilde W}_{n}=-4\sum_{k,l=1}^{\infty}a_{-k}a_{-l}a_{k+l+n}
-2\sum_{k=1}^{\infty}\sum_{l=0}^{k+n}a_{-k}a_la_{k-l+n}
+2Q\sum_{k=0}^{\infty}(n+k+1)a_{-k}a_{k+n}.
\end{equation}
In similarity with (\ref{Mncon}), we have
\begin{eqnarray}\label{DMncon}
&&\prod_{i=1}^N\int_{-\infty}^{+\infty} dz_i\langle U_{N\sqrt{\beta},t}|\mathcal{\tilde W}_{n}
\prod_{i=1}^{N}V^{-\sqrt{\beta}}(z_i)|0\rangle\nonumber\\
&=&\prod_{i=1}^N\int_{-\infty}^{+\infty} dz_i{\bar {\mathcal{W}}}_{n}\langle U_{N\sqrt{\beta},t}|
\prod_{i=1}^{N}V^{-\sqrt{\beta}}(z_i)|0\rangle=0,
\end{eqnarray}
where the operators $\bar {\mathcal{W}}_{n}$ are given by
\begin{equation}\label{barMn}
\bar {\mathcal{W}}_n=\sum_{i=1}^{N}\frac{\partial^2}{\partial z^2_i}z^{n+2}_i
-\gamma\sum_{i=1}^{N}\sum_{j\neq i}\frac{\partial}{\partial z_i}\frac{z^{n+2}_i}{z_i-z_j}-(n+2)
\sum_{i=1}^{N}\frac{\partial}{\partial z_i}z^{n+1}_i, \ \ n\geq -2,
\end{equation}
and $\gamma=2\beta$.

We obtain the constraints from (\ref{DMncon})
\begin{equation}\label{DMntcon}
\hat {\mathcal{W}}_{n}Z_{\beta}=0,\ \ n\geq -2,
\end{equation}
where
\begin{eqnarray}\label{DMnt}
\hat {\mathcal{W}}_{n}&=&\sum_{k,l=1}^{\infty}klt_kt_l\frac{\partial}{\partial t_{k+l+n}}
+\beta\sum_{k=1}^{\infty}\sum_{l=0}^{k+n}kt_k\frac{\partial}{\partial t_l}\frac{\partial}{\partial t_{k-l+n}}\nonumber\\
&&+(1-\beta)\sum_{k=0}^{\infty}kt_k(n+k+1)\frac{\partial}{\partial t_{k+n}}.
\end{eqnarray}

In similarity with the case of hermitian one-matrix model,
we may construct the  Virasoro constraints
\begin{equation}\label{dWcon}
{\tilde {\mathbb{L}}}_mZ_{\beta}=0,
\end{equation}
with ${\tilde {\mathbb{L}}}_m=\frac{1}{n}\hat {\mathcal{W}}_{n}^m\hat {\mathcal{L}}_0,$ $m\in \mathbb{N}$, for any given $n$.
The constraint operators
${\tilde {\mathbb{L}}}_m$ yield the centerless Virasoro algebra  and null 3-algebra (\ref{nullalg}).

Similarly to (\ref{genvir}), we may introduce the operators
\begin{equation}\label{Dgenvir}
{\mathcal{L}}^m_{k_1,\cdots,k_r}=a_{k_1}\cdots a_{k_r}{\mathcal{L}}_m,
 \ \ m\geq -1, k_1,\cdots,k_r\geq 0,
\end{equation}
such that they yield the generalized centerless Virasoro algebra (\ref{genviralg}) and
null 3-algebra (\ref{genvirnullalg}). Inserting (\ref{Dgenvir}) into the expectation value in (\ref{DPF}),
we may derive the constraints
\begin{equation}
\hat {\mathcal{L}}^m_{k_1,\cdots,k_r}Z_{\beta}=0,\ \ m\geq -1,
\end{equation}
where
\begin{eqnarray}
\hat {\mathcal{L}}^m_{k_1,\cdots,k_r}&=&\sum_{k=1}^{\infty}kt_k\frac{\partial}{\partial t_{k+m}}
\frac{\partial}{\partial t_{k_1}}\cdots \frac{\partial}{\partial t_{k_r}}+\sum_{i=1}^{r}k_i
\frac{\partial}{\partial t_{k_1}}\cdots  \frac{\partial}{\partial t_{k_{i-1}}}\frac{\partial}{\partial t_{m+k_i}}
\frac{\partial}{\partial t_{k_{i+1}}}\cdots \frac{\partial}{\partial t_{k_r}}
\nonumber\\
&&+\beta\sum_{k=0}^{m}\frac{\partial}{\partial t_{k}}
\frac{\partial}{\partial t_{m-k}}\frac{\partial}{\partial t_{k_1}}\cdots \frac{\partial}{\partial t_{k_r}}
+(1-\beta)(m+1)\frac{\partial}{\partial t_{m}}\frac{\partial}{\partial t_{k_1}}\cdots \frac{\partial}{\partial t_{k_r}},
\end{eqnarray}
and they yield the generalized centerless Virasoro algebra (\ref{genviralg}) and null 3-algebra (\ref{genvirnullalg}) as well.

\subsection{Total derivative operators $\bar {\mathcal{W}}_n$ and extended operators $\mathscr{H}_n$}
\subsubsection{Total derivative operators $\bar {\mathcal{W}}_n$}

We have derived the total derivative operators $\bar {\mathcal{W}}_n$ (\ref{barMn}) associating with the
constraints (\ref{DMntcon}). In the following, we focus on analyzing these operators.

Let us take $t_k=-\frac{1}{2}\delta_{k,2}$ and $\gamma=2\beta$ in (\ref{DPF}), then the integrand
becomes the real Laughlin wave function
\begin{equation}\label{vandegauss}
\Phi(z)=\langle U_{N\sqrt{\frac{\gamma}{2}},t}|\prod_{i=1}^{N}V^{-\sqrt{\frac{\gamma}{2}}}(z_i)|0\rangle
=\Delta(z)^{\gamma}{\rm exp}(-\frac{1}{2}\sum_{i=1}^{N}z_i^2).
\end{equation}

We recognize (\ref{vandegauss}) to be the ground state eigenfunction for the $A$-type Calogero model which is a one-dimensional
integrable system with inverse-square long-range interaction confined in an external harmonic well of
strength $\omega=1$ \cite{Calogero1971,SutherlandJMP}
\begin{equation}\label{Calogero}
H^A_{C}=\sum_{i=1}^{N}(-\frac{\partial^2}{\partial z_i^2}+z_i^2)+V^A_{C},
\end{equation}
where the potential $V^A_C$ is given by
\begin{equation}\label{VC}
V^A_C=\gamma(\gamma-1)\sum_{j\neq i}\frac{1}{(z_i-z_j)^2}.
\end{equation}
The ground state energy of (\ref{Calogero}) is $E^C_0=N(1+\gamma (N-1))$.

Let us consider the actions of $\bar {\mathcal{W}}_n$  on the wave function (\ref{vandegauss}).
It is not difficult to obtain the constraints
\begin{equation}
(\bar {\mathcal W}_{n-2}-\bar L_n)\Phi(z)=0,\quad n\geq 0,
\end{equation}
where the operators $\bar L_n=\sum_{i=1}^{N}\frac{\partial}{\partial z_i}z_i^{n+1}$
are the generators  of the Virasoro algebra.
When particularized to the operators $\bar {\mathcal W}_{-2}$ and $\bar L_0$, we may construct the generators
${\bar {\mathbb{L}}}_m=\frac{1}{2}\bar {\mathcal{W}}_{-2}^m(\bar{\mathcal{W}}_{-2}-\bar L_0),$
 $m\in \mathbb{N}$ which yield the centerless Virasoro algebra and the null $3$-algebra (\ref{nullalg}).
It is clear that the real Laughlin wave function (\ref{vandegauss}) is annihilated by the operators ${\bar {\mathbb{L}}}_m$.

Under the similarity transformation of the wave function (\ref{vandegauss})
and removing the ground state from the Hamiltonian (\ref{Calogero}), we obtain
\begin{eqnarray}
\Phi(z)^{-1}(H^A_C-E^C_0)\Phi(z)=-O^A_{L},
\end{eqnarray}
where $O_L^A$ is the $A$-type
Lassalle operator \cite{Lassalle1}
\begin{eqnarray}\label{LA}
O_L^A=\sum_{i=1}^{n}\frac{\partial^2}{\partial z_i^2}
+\frac{2}{\alpha}\sum_{i\neq j}\frac{1}{z_i-z_j}\frac{\partial}{\partial z_j}
-2\sum_{i=1}^{n}z_i\frac{\partial}{\partial z_i}.
\end{eqnarray}
Here we have taken the strength of interactions $\gamma=\frac{1}{\alpha}$ in (\ref{VC}) for convenience.

In terms of the $A$-type Lassalle operator $O_L^A$, Euler operator $O_E=\sum_{i=1}^{N}z_i\frac{\partial}{\partial z_i}$
and potential $V^A_C$, the total derivative operator $\bar {\mathcal W}_{-2}$ can be expressed as
\begin{eqnarray}\label{ew2}
\bar {\mathcal W}_{-2}=O_L^A+2O_E+\frac{2\alpha}{\alpha-1}V_C^A.
\end{eqnarray}

Similarly, the total derivative operator $\bar {\mathcal W}_{-1}$ can be expressed as
\begin{eqnarray}\label{ew1}
\bar {\mathcal W}_{-1}=O_L^B+O_E+\frac{2\alpha}{\alpha-1}V_C^B,
\end{eqnarray}
where $O_L^B$ is the $B$-type Lassalle operator \cite{Lassalle2}
\begin{equation}\label{LB}
O_L^B=\sum_{i=1}^{N}z_i\frac{\partial^2 }{\partial z_i^2}
+\frac{2}{\alpha}\sum_{i\neq j}\frac{z_i}{z_i-z_j}\frac{\partial }{\partial z_i}
+(a+1)\sum_{i=1}^{N}\frac{\partial }{\partial z_i}
-\sum_{i=1}^{N}z_i\frac{\partial }{\partial z_i},
\end{equation}
with $a=0$ in (\ref{ew1}),
$V^B_C$ is the potential of $B$-type Calogero model \cite{Olshanetsky}, i.e.,
$V^B_C=\frac{1}{\alpha}(\frac{1}{\alpha}-1)\sum_{j\neq i}\frac{z_i}{(z_i-z_j)^2}$.

Taking $t=0$ and $2\beta=\gamma$ in the integrand of (\ref{DPF}),  we have
\begin{equation}\label{vande}
\langle U_{N\sqrt{\frac{\gamma}{2}},0}|\prod_{i=1}^{N}V^{-\sqrt{\frac{\gamma}{2}}}(z_i)|0\rangle
=\Delta(z)^{\gamma},
\end{equation}
which is the ground state eigenfunction for the many body problem \cite{Sogo}
\begin{equation}\label{Le}
H_{L}=\sum_{i=1}^{N}((-1+z_i^2)\frac{\partial^2}{\partial z_i^2}+2z_i\frac{\partial}{\partial z_i}
+\gamma(\gamma-1)\sum_{j\neq i}\frac{1-z_i^2}{(z_i-z_j)^2}),
\end{equation}
and the ground state energy is $E^L_0=\gamma N(N-1)+\frac{1}{3}\gamma^2N(N-1)(N-2)$.

In addition, it is noted that the ground state eigenfunction of the Sutherland model \cite{Sutherland1971}
\begin{eqnarray}\label{Sutherland model}
H_{S}=-\sum_{i=1}^{N}\frac{\partial^2}{\partial \theta_i^2}
+\frac{1}{2}\gamma(\gamma-1)\sum_{1\leq i<j\leq N}\frac{1}{\mathrm{sin}^2((\theta_i-\theta_j)/2)}
\end{eqnarray}
is proportional to (\ref{vande}).
More precisely, the Hamiltonian (\ref{Sutherland model}) can be expressed as
\begin{eqnarray}\label{Sutherland model1}
H_{S}=\sum_{i=1}^{N}(z_i\frac{\partial}{\partial z_i})^2
+V_S,
\end{eqnarray}
where $z_i=e^{\mathbf{i}\theta_j}$, $\mathbf{i}=\sqrt{-1}$,
\begin{equation}\label{VS}
V_S=-2\gamma(\gamma-1)\sum_{1\leq i<j\leq N}\frac{z_iz_j}{(z_i-z_j)^2}.
\end{equation}
The ground state eigenfunction and energy of (\ref{Sutherland model1}) are
$\psi=\frac{1}{\sqrt{N!}}(\prod_{j=1}^{N}z_j)^{-\gamma\frac{N-1}{2}}
\Delta(z)^{\gamma}$
and $E^S_0=\frac{1}{12}\gamma^2N(N^2-1)$, respectively.

When $\gamma$ is an even integer, $(\ref{vande})$ is a symmetric function. In this case, $\Delta(z)^{\gamma}$
can be expanded as the linear combination of the Schur functions. The calculations of the coefficients in the
expansion become computationally expensive as $\gamma$ increases. Several algorithms have been discussed
in Refs. \cite{Scharff}-\cite{CBallantine}, but no general formulas are given yet. It is interesting to note that
the actions of $\bar {\mathcal{W}}_{n}$ on the state (\ref{vande}) give the constraints
\begin{equation}\label{cvander}
\bar {\mathcal{W}}_{n}\Delta(z)^{\gamma}=0.
\end{equation}

For the Sutherland model $(\ref{Sutherland model1})$,
there is the similarity transformation
\begin{equation}\label{simtrans}
{\psi}^{-1}(H_{S}- E^S_0)\psi=H_{LB},
\end{equation}
where $H_{LB}$ is the Laplace-Beltrami operator
\begin{eqnarray}\label{LBoperator}
H_{LB}=\sum_{i=1}^{N}(z_i\frac{\partial }{\partial z_i})^2+\frac{1}{\alpha}
\sum_{i<j}\frac{z_i+z_j}{z_i-z_j}(z_i\frac{\partial }{\partial z_i}-z_j\frac{\partial }{\partial z_j}).
\end{eqnarray}
Here we have also taken $\gamma=\frac{1}{\alpha}$ in (\ref{VS}) for convenience.

In terms of the Laplace-Beltrami operator $H_{LB}$, Euler operator $O_E$ and potential $V_S$,
the total derivative operator $\bar {\mathcal{W}}_{0}$ can be expressed as
\begin{equation}\label{ew0}
\bar {\mathcal{W}}_{0}=H_{LB}+(1+\frac{N-1}{\alpha})O_E-
\frac{2\alpha}{\alpha-1}V_S+\frac{1}{\alpha}N(N-1).
\end{equation}

\subsubsection{Extended operators $\mathscr{H}_n$ and eigenfunctions}
Recall that the Hamiltonian of the Sutherland model with the exchange operator  is given by \cite{Polychronakos}
\begin{equation}
\bar H_{S}=-\sum_{i=1}^{N}\frac{\partial^2}{\partial \theta_i^2}
+\frac{1}{2\alpha}\sum_{1\leq i<j\leq N}\frac{1/\alpha-s_{ij}}{\mathrm{sin}^2((\theta_i-\theta_j)/2)},
\end{equation}
where $s_{ij}$ is a coordinate exchange operator which is defined by the action
$s_{ij}f(z_1,\cdots,z_i,\cdots,z_j,$\\
$\cdots,z_N)=f(z_1,\cdots,z_j,
\cdots,z_i,\cdots,z_N)$.
It has the same ground state eigenfunction as (\ref{Sutherland model1}).
Under the similarity transformation (\ref{simtrans}),
we obtain the Laplace-Beltrami operator with the exchange operator
\begin{eqnarray}\label{LBEO}
\mathscr{H}_{LB}=\sum_{i=1}^{N}(z_i\frac{\partial }{\partial z_i})^2+\frac{N-1}{\alpha}\sum_{i=1}^{N}z_i\frac{\partial }
{\partial z_i}+\frac{2}{\alpha}\sum_{i<j}\frac{z_iz_j}{z_i-z_j}(\frac{\partial }{\partial z_i}-\frac{\partial }
{\partial z_j}-\frac{1-s_{ij}}{z_i-z_j}).
\end{eqnarray}
When $s_{ij}=1$, (\ref{LBEO}) gives the Laplace-Beltrami operator (\ref{LBoperator}).

Let us construct the extended operators with the exchange operator $s_{ij}$
\begin{eqnarray}\label{GHS}
\mathscr{H}_n={\mathscr{\bar W}}_{n}+\kappa O_E,\quad \kappa\neq 0,
\end{eqnarray}
where
the operators ${\mathscr{\bar W}}_{n}$ are given by
\begin{eqnarray}\label{GMS}
{\mathscr{\bar W}}_{n}&=&\sum_{i=1}^{N}z_i^{n+2}\frac{\partial^2 }{\partial z_i^2}
+\frac{2}{\alpha}\sum_{i\neq j}\frac{z^{n+2}_i}{z_i-z_j}\frac{\partial }{\partial z_i}
+\zeta\sum_{i=1}^{N}z_i^{n+1}\frac{\partial }{\partial z_i}\nonumber\\
&+&\sum_{i\neq j}(\vartheta
\frac{z^{n+1}_i}{z_i-z_j}-\frac{1}{\alpha}
\frac{z^{n+2}_i}{(z_i-z_j)^2})(1-s_{ij}),
\end{eqnarray}
and the commutator of ${\mathscr{\bar W}}_n$ and the Euler operator is
$[{\mathscr{\bar W}}_n, O_E]=-n{ \mathscr{\bar W}}_n.$

Note that the operators (\ref{GMS}) are the extended case of $\bar {\mathcal{W}}_n$ (\ref{barMn}).
When we take $\alpha=-\frac{2}{\gamma}$, $\vartheta=-\frac{\gamma}{2}(n+2)$, $\zeta=n+2$ and $s_{ij}=-1$ in
(\ref{GMS}), it reduces to  $\bar {\mathcal{W}}_n$ (\ref{barMn}).

The operator $\mathscr{H}_{LB}$ (\ref{LBEO}) is a particular case of $\mathscr{H}_0$.
In terms of the mutually commuting Cherednik operators \cite{Cherednik}
\begin{equation}
\xi_i=\alpha z_i(\frac{\partial }{\partial z_i}+\frac{1}{\alpha}\sum_{j\neq i}
\frac{1-s_{ij}}{z_i-z_j})+1-N+\sum_{j=i+1}^N s_{ij},\quad i=1,\cdots, N,
\end{equation}
there is the expression
\begin{eqnarray}
\mathscr{H}_{LB}
=\frac{1}{\alpha^2}\sum_{i=1}^{N}(\xi_i+\frac{N-1}{2})^2-E^S_0.
\end{eqnarray}

The Cherednik operators $\xi_i$ have the nonsymmetric Jack polynomials
$\bar J_{\eta}(z;\alpha)=z^{\eta}+\sum_{\rho \prec \eta}b_{\eta \rho}z^{\rho}$,
$b_{\eta \rho} \in \mathbb{Q}(\alpha)$ \cite{Opdam} as the eigenfunctions, i.e.,
\begin{equation}
\xi_i \bar J_{\eta}(z;\alpha)=\bar{\eta}_i \bar J_{\eta}(z;\alpha),
\end{equation}
where $\eta$ is a composition $\eta=(\eta_1,\cdots, \eta_N)$, $\eta_i\in \mathbb{N}$,
$z^{\eta}=z_1^{\eta_{1}}\cdots z_N^{\eta_{N}}$,
$\bar{\eta}_i=\alpha{\eta}_i-\sharp \{j<i|\eta_j\geq \eta_i \}-\sharp \{j>i|\eta_j>\eta_i \}$.
Therefore, $\mathscr{H}_{LB}$ has the nonsymmetric Jack polynomials as the eigenfunctions.
It is noted that the symmetric Jack polynomials are also the eigenfunctions of $\mathscr{H}_{LB}$.
However, the operator $\mathscr{H}_0$ only has the symmetric Jack polynomials as its eigenfunctions.

The Euler operator can be transformed into the Hamiltonian of
the decoupled quantum harmonic oscillators \cite{Nishino}
\begin{equation}
{\rm{exp}}({-\frac{\kappa}{2}\sum_{i=1}^Nz_i^2})
{\rm{exp}}({-\frac{1}{4\kappa}\sum_{i=1}^N\frac{\partial^2 }{\partial z_i^2}})
 O_E{\rm{exp}}({\frac{1}{4\kappa}\sum_{i=1}^N\frac{\partial^2 }{\partial z_i^2}})
{\rm{exp}}({\frac{\kappa}{2}\sum_{i=1}^Nz_i^2})=\sum_{j=1}^{N}A_j^{\dagger}A_j,
\end{equation}
where $A_j=\frac{\partial }{\partial z_j}+\kappa z_j$ and
$A_j^{\dagger}=\frac{1}{2\kappa}(-\frac{\partial }{\partial z_j}+\kappa z_j)$, $j=1,\cdots, N$.
Thus we have
\begin{equation}\label{transfrelation}
\mathscr{T}^{-1}\mathscr{H}_n\mathscr{T}=\kappa\sum_{j=1}^{N}A_j^{\dagger}A_j,\quad n\neq 0,
\end{equation}
where the transformation operators $\mathscr{T}$ are given by
\begin{equation}
\mathscr{T}={\rm{exp}}({-\frac{1}{n\kappa}{ \mathscr{\bar W}}_n})
{\rm{exp}}({\frac{1}{4\kappa}\sum_{i=1}^N\frac{\partial^2 }{\partial z_i^2}})
{\rm{exp}}({\frac{\kappa}{2}\sum_{i=1}^Nz_i^2}).
\end{equation}

The number operator $\sum_{j=1}^{N}A_j^{\dagger}A_j$ has the eigenstates
\begin{equation}\label{neigenstates}
|\eta_{1},\cdots,\eta_{N}\rangle=
\prod_{j=1}^{N}(A_j^{\dagger})^{\eta_j}|0\rangle,
\end{equation}
where $|0\rangle={\rm{exp}}({-\frac{\kappa}{2}\sum_{i=1}^Nz_i^2})$.
It is easy to know that
\begin{eqnarray}\label{nonsymteigen}
\mathscr{T}|\eta_1,\cdots,\eta_N\rangle
={\rm{exp}}(-\frac{1}{n\kappa}{ \mathscr{\bar W}}_n)
z^{\eta}
\end{eqnarray}
are the eigenstates of $\mathscr{H}_n$ ($n\neq 0$)
\begin{equation}\label{eigeneqn}
\mathscr{H}_n\mathscr{T}|\eta_1,\cdots,\eta_N\rangle
=\kappa|\eta|\mathscr{T}|\eta_1,\cdots,\eta_N\rangle,
\end{equation}
in which $|\eta|=\sum_{i=1}^{N}\eta_i$. The eigenstates (\ref{nonsymteigen}) have no singularities
except for $z_i=0$, $i=1,\cdots, N$.

Actually, there are more general eigenfunctions of $\mathscr{H}_n$ ($n\neq 0$).
Let us assume that $f$ is a homogeneous polynomial with degree $|f|$.
Define the functions
\begin{equation}\label{eigenfunction}
\psi_n={\rm{exp}}({-\frac{1}{n\kappa}{ \mathscr{\bar W}}_n})f,\quad n\neq 0,
\end{equation}
it is not difficult to see that
$\psi_n$ are also the eigenfunctions of $\mathscr{H}_n$
\begin{eqnarray}
\mathscr{H}_n\psi_n=\kappa |f| \psi_n, \quad n\neq 0.
\end{eqnarray}

Taking  $\zeta=0$, $\vartheta=0$,  $\kappa=-2$, $n=-2$ in (\ref{GHS}),
it gives the $A$-type Lassalle operator with the exchange operator
\begin{equation}
{\bar O}_{L}^{A}=\sum_{i=1}^{n}\frac{\partial^2}{\partial z_i^2}
+\frac{2}{\alpha}\sum_{i\neq j}\frac{1}{z_i-z_j}\frac{\partial}{\partial z_j}
-\frac{1}{\alpha}\sum_{i\neq j}
\frac{1}{(z_i-z_j)^2}(1-s_{ij})
-2\sum_{i=1}^{n}z_i\frac{\partial}{\partial z_i},
\end{equation}
which has the generalized nonsymmetric Hermite polynomials
\begin{equation}\label{nonherpoly}
\bar{J}^{H}_{\eta}(z;\alpha)=\bar{J}_{\eta}(z;\alpha)+\sum_{|\rho|<|\eta|}b^H_{\eta\rho}\bar{J}_{\rho}(z;\alpha),\ \
b^H_{\eta\rho}\in \mathbb{Q}(\alpha)
\end{equation}
as the eigenfunctions \cite{Baker1997}. It is noted that the eigenfunctions (\ref{nonherpoly})
are the special cases of (\ref{eigenfunction}) with $f=\bar{J}_{\eta}(z;\alpha)$.

Taking $\zeta=a+1$, $\vartheta=0$, $\kappa=-1$, $n=-1$ in (\ref{GHS}), it gives the $B$-type Lassalle operator
with the exchange operator
\begin{eqnarray}
{\bar O}_{L}^{B}&=&\sum_{i=1}^{N}z_i\frac{\partial^2 }{\partial z_i^2}
+\frac{2}{\alpha}\sum_{i\neq j}\frac{z_i}{z_i-z_j}\frac{\partial }{\partial z_i}+(a+1)\sum_{i=1}^{N}\frac{\partial }
{\partial z_i}\nonumber\\
&-&\frac{1}{\alpha}\sum_{i\neq j}\frac{z_i}{(z_i-z_j)^2}(1-s_{ij})
-\sum_{i=1}^{N}z_i\frac{\partial }{\partial z_i},
\end{eqnarray}
which has the generalized nonsymmetric Laguerre polynomials
\begin{equation}\label{nonlagfun}
\bar{J}^{L}_{\eta}(z;\alpha)=\bar{J}_{\eta}(z;\alpha)
+\sum_{|\rho|<|\eta|}b^L_{\eta\rho}\bar{J}_{\rho}(z;\alpha),
\ \ b^L_{\eta\rho}\in \mathbb{Q}(\alpha,a)
\end{equation}
as the eigenfunctions \cite{Baker1997}. When $f=\bar{J}_{\eta}(z;\alpha)$ in (\ref{eigenfunction}),
we have the eigenfunctions (\ref{nonlagfun}).

Let us turn to the case  $s_{ij}=1$ in (\ref{GHS}), i.e.,
\begin{eqnarray}\label{GH}
H_n=W_{n}+\kappa O_E,\quad \kappa\neq 0,
\end{eqnarray}
where $W_{n}$ are given by
\begin{eqnarray}\label{GM}
W_{n}=\sum_{i=1}^{N}z_i^{n+2}\frac{\partial^2 }{\partial z_i^2}
+\frac{2}{\alpha}\sum_{i\neq j}\frac{z^{n+2}_i}{z_i-z_j}\frac{\partial }{\partial z_i}
+\zeta\sum_{i=1}^{N}z_i^{n+1}\frac{\partial }{\partial z_i}.
\end{eqnarray}

The operator $H_0$ has the Jack symmetric polynomials $J_{\lambda}(z;\alpha)$ as the eigenfunctions.
In particular, for the case of $\zeta+\kappa=1-\frac{1}{\alpha}(N-1)$ in $H_0$,
we have the Laplace-Beltrami operator (\ref{LBoperator}).

For the operators $H_n$ ($n\neq 0$), in order to obtain the eigenfunctions which have no singularities
except for $z_i=0$, $i=1,\cdots, N$, we need to define the symmetrized states associated with the partition
$\lambda=(\lambda_1,\cdots,\lambda_N)$ \cite{Nishino}
\begin{equation}
|\lambda\rangle=\sum_{\sigma \in S_N}|\lambda_{\sigma(1)},\cdots,\lambda_{\sigma(N)}\rangle.
\end{equation}
Then
\begin{eqnarray}\label{teigen}
T|\lambda\rangle={\rm{exp}}(-\frac{1}{n\kappa}W_n)m_{\lambda}
\end{eqnarray}
have no singularities except for $z_i=0$, $i=1,\cdots, N$, where
\begin{equation}
T={\rm{exp}}({-\frac{1}{n\kappa}W_n})
{\rm{exp}}({\frac{1}{4\kappa}\sum_{i=1}^N\frac{\partial^2 }{\partial z_i^2}})
{\rm{exp}}({\frac{\kappa}{2}\sum_{i=1}^Nz_i^2}),
\end{equation}
and $m_\lambda=\sum_{\sigma \in S_N}z_1^{\lambda_{\sigma(1)}}\cdots z_N^{\lambda_{\sigma(N)}}$.

It is easy to see that (\ref{teigen}) are the eigenfunctions of $H_n$. Similar to the operators $\mathscr{H}_n$
($n\neq 0$),
$H_n$ ($n\neq 0$) have the eigenfunctions which possess the form (\ref{eigenfunction}).
Let us take $f=J_{\lambda}(z;\alpha)$, $s_{ij}=1$ in (\ref{eigenfunction}) and define
\begin{equation}\label{symeigenfunction}
\varphi_{n,\lambda}={\rm {exp}}(-\frac{W_n}{\kappa n})J_{\lambda}(z;\alpha),\quad n\neq 0.
\end{equation}
These  eigenfunctions have close relations with the generalized Hermite (Laguerre)
polynomials \cite{Lassalle1,Lassalle2},\cite{Macdonaldpaper,BakerC}.
It is clear that the  $A$ and $B$-type Lassalle operators (\ref{LA}) and (\ref{LB})
are the particular cases of (\ref{GH}).
Their eigenfunctions are the generalized Hermite polynomials
$\varphi_{-2,\lambda}=J_{\lambda}+\sum_{\mu\subset \lambda}c_{\lambda\mu}^{H}J_{\mu}$, $c_{\lambda\mu}^{H}\in \mathbb{Q}(\alpha)$
and the generalized Laguerre polynomials $\varphi_{-1,\lambda}=J_{\lambda}+\sum_{\mu\subset \lambda}c_{\lambda\mu}^{L}J_{\mu}$,
$c_{\lambda\mu}^{L}\in \mathbb{Q}(\alpha,a)$, respectively.

\subsection{Superintegrability of the $\beta$-deformed $N\times N$ complex matrix model}

We have obtained the constraints (\ref{DMntcon}) for the $\beta$-deformed hermitian matrix model.
To explore the intrinsic connection between the derived constraint operators (\ref{DMnt}) and
$W$-representations of some matrix models,
let us take the rescaling variables $p_k=\beta^{-1}kt_k$ $(k>0)$, and substitute
$\frac{\partial}{\partial t_0}$ by $N$  in (\ref{DMnt}).
We then have
\begin{eqnarray}\label{dhMnp}
\hat {\mathcal{W}}_{n}&=&\sum_{k,l=1}^{\infty}\big(\beta(k+l+n)p_{k}p_l
\frac{\partial}{\partial p_{k+l+n}}+klp_{k+l-n}\frac{\partial}{\partial p_k}\frac{\partial}{\partial p_l}\big)\nonumber\\
&&+2\beta N\sum_{k=1}^{\infty}(k+n)p_{k}
\frac{\partial}{\partial p_{k+n}}+(1-\beta)\sum_{k=1}^{\infty}(n+k+1)(k+n)p_k\frac{\partial}{\partial p_{k+n}}
\nonumber\\
&&+\beta N[\beta p_1^2+(\beta N+1-\beta)p_2]\delta_{n,-2}
+\beta N(\beta N+1-\beta)p_1\delta_{n,-1}, \quad n\geq -2.
\end{eqnarray}

For the operators $\hat {\mathcal{W}}_{-2}$ and $\hat {\mathcal{W}}_{-1}$, they appear in
the $W$-representations of $\beta$-deformed Gaussian hermitian \cite{Morozov1901} and $N\times N$
complex matrix models \cite{Cheny,Cassia2020}, respectively,
\begin{eqnarray}
{\mathcal Z}_G\{p\}&=&\prod_{i=1}^N\int_{-\infty}^{+\infty} dz_i\Delta(z)^{2\beta}
e^{\sum_{i=1}^{N}\sum_{k=1}^{\infty}\beta\frac{ p_k}{k}z_i^k-
\frac{1}{2}\sum_{i=1}^Nz_i^2}
=e^{{\hat {\mathcal{W}}}_{-2}/2}\cdot 1,\label{dgm}\\
{\mathcal Z}_C\{p\}&=&\prod_{i=1}^N\int_{0}^{+\infty} dz_i \Delta(z)^{2\beta}
e^{\sum_{i=1}^{N}\sum_{k=1}^{\infty}\beta\frac{p_k}{k}z_i^k-
\frac{1}{2}\sum_{i=1}^Nz_i}
=e^{{\hat {\mathcal{W}}}_{-1}}\cdot 1.\label{dcm}
\end{eqnarray}

The $\beta$-deformed Gaussian hermitian matrix model can be expanded in terms of the Jack symmetric
polynomials $J_{\lambda}\{p\}$ as follows \cite{Morozov1901}
\begin{eqnarray}
{\mathcal Z}_G\{p\}=\sum_{\lambda}\beta^{|\lambda|/2}
\frac{J_{\lambda}\{p_k=\delta_{k,2}\}J_{\lambda}\{p_k=N\}}{J_{\lambda}\{p_k=\delta_{k,1}\}
\langle J_{\lambda}, J_{\lambda}\rangle_{\beta}}
J_{\lambda}\{p\}.
\end{eqnarray}
Here the scalar product $\langle \ .\  , \ .\ \rangle_{\beta}$ is given by
\begin{equation}
\langle p_{\lambda}, p_{\mu}\rangle_{\beta}=\delta_{\lambda,\mu}z_{\lambda}\beta^{-l(\lambda)}.
\end{equation}
$\langle J_{\lambda}, J_{\lambda}\rangle_{\beta}=\frac{h_{\lambda}}{h^{'}_{\lambda}}$,
$h_{\lambda}=\prod_{(i,j)\in \lambda}(1+\lambda_i-j+\beta(\lambda_j^{'}-i))$ and
$h^{'}_{\lambda}=\prod_{(i,j)\in \lambda}(\lambda_i-j+\beta(\lambda_j^{'}-i+1))$ are the deformed hook length,
in which $\lambda^{'}=(\lambda_1^{'},\lambda_2^{'},\cdots)$ is the conjugate partition of $\lambda$.

In the following, let us derive the character expansion of (\ref{dcm}) from the $W$-representation.
It is known that the Jack polynomials are the eigenfunctions of the operator \cite{Jing}
\begin{eqnarray}
\hat {\mathscr{W}}_0=\frac{1}{2}\sum_{k,l=1}^{\infty}\big(\beta(k+l)p_{k}p_l
\frac{\partial}{\partial p_{k+l}}+klp_{k+l}\frac{\partial}{\partial p_k}\frac{\partial}{\partial p_l}\big)
+\frac{1}{2}(1-\beta)\sum_{k=1}^{\infty}k^2p_k\frac{\partial}{\partial p_{k}},
\end{eqnarray}
and
\begin{equation}{\label{defaction0}}
\hat {\mathscr{W}}_0 J_{\lambda}=\frac{1}{2}\sum_{i=1}^{l(\lambda)}\lambda_i(\lambda_i+\beta(-2i+1))J_{\lambda}.
\end{equation}
The action of $p_1$ on the Jack polynomials is given by
\begin{equation}{\label{defactionp1}}
p_1J_{\lambda}=\sum_{\lambda+\square}\dbinom{\lambda+\square}{\lambda}\frac{h_{\lambda}}{h_{\lambda+\square}}
J_{\lambda+\square},
\end{equation}
where $\dbinom{\lambda+\square}{\lambda}$ are the generalized binomial coefficients defined by \cite{Lassalle1998}
\begin{equation}
\dbinom{\lambda+\square}{\lambda}=(\beta l(\lambda+\square)+j_{\square}-\beta i_{\square})
\prod_{j\neq i_{\square}}\frac{j_{\square}-\lambda_j+\beta(j-i_{\square}-1)}{j_{\square}
-\lambda_j+\beta(j-{i_{\square}})}.
\end{equation}

In terms of $\hat{\mathscr{W}_0}$ and  $p_{1}$, we may rewrite the operator ${\hat {\mathcal{W}}}_{-1}$ as
\begin{eqnarray}
{\hat {\mathcal{W}}}_{-1}=[\hat{\mathscr{W}_0},[\hat{\mathscr{W}_0}, p_{1}]]+2N\beta[\hat{\mathscr{W}_0}, p_{1}]
+(\beta^2 N^2-\frac{1}{4}(\beta-1)^2)p_{1},
\end{eqnarray}
and give the action
\begin{eqnarray}{\label{defactionw1}}
{\hat {\mathcal{W}}}_{-1}J_{\lambda}=\beta^2\sum_{\lambda+\square} c(i_\square,j_\square)
(c(i_\square,j_\square)+\beta^{-1}-1)\dbinom{\lambda+\square}{\lambda}\frac{h_{\lambda}}{h_{\lambda+\square}}
J_{\lambda+\square},
\end{eqnarray}
where $c(i_\square,j_\square)=\beta^{-1}(j_\square-1)-(i_\square-1)+N$.

Then it is straightforward to calculate the power of ${\hat {\mathcal{W}}}_{-1}$ acting on
$J_{\lambda}$ with $\lambda=\varnothing$,  leading to the explicit result
\begin{eqnarray}\label{defactionw1m}
{\hat {\mathcal{W}}}^m_{-1}\cdot 1&=&\beta^{2m}\sum_{\lambda\mapsto m}\prod_{(i,j)\in \lambda} c(i,j)
(c(i,j)+\beta^{-1}-1)b(\lambda)J_{\lambda}\nonumber\\
&=&\sum_{\lambda\mapsto m}\frac{J_{\lambda}\{p_k=N\}J_{\lambda}
\{p_k=N+\beta^{-1}-1\}}{J_{\lambda}\{p_k=\beta^{-1}\delta_{k,1}\}
\langle J_{\lambda}, J_{\lambda}\rangle_{\beta}}
J_{\lambda}\{p\},
\end{eqnarray}
where
\begin{equation}
b(\lambda)=\frac{\langle p_1^m,J_{\lambda}\rangle_{\beta}}{\langle J_{\lambda}, J_{\lambda}\rangle_{\beta}}
=m!\beta^{-m}\frac{J_{\lambda}\{p_k=\delta_{k,1}\}}{\langle J_{\lambda}, J_{\lambda}\rangle_{\beta}},
\end{equation}
and we have used the formula
\begin{equation}
\frac{J_{\lambda}\{p_k=N\}}{J_{\lambda}\{p_k=\delta_{k,1}\}}=\prod_{(i,j)\in \lambda}c(i,j).
\end{equation}

Thus we finally reach the desired result
\begin{equation}\label{expaninjack}
{\mathcal Z}_C\{p\}=e^{{\hat {\mathcal{W}}}_{-1}}\cdot 1=\sum_{\lambda}\frac{J_{\lambda}\{p_k=N\}
J_{\lambda}\{p_k=N+\beta^{-1}-1\}}{J_{\lambda}\{p_k=\beta^{-1}\delta_{k,1}\}
\langle J_{\lambda}, J_{\lambda}\rangle_{\beta}}J_{\lambda}\{p\}.
\end{equation}
In addition, using the Cauchy formula
\begin{equation}
\sum_{\lambda}\frac{1}{\langle J_{\lambda}, J_{\lambda}\rangle_{\beta}}J_{\lambda}\{p_k\}
J_{\lambda}\{\bar p_k\}=e^{\beta\sum_{k=1}^{\infty}\frac{p_k\bar p_k}{k}},
\end{equation}
the partition function ${\mathcal Z}_C\{p\}$ can be expanded as
\begin{equation}\label{expaninjack1}
{\mathcal Z}_C\{p\}=\sum_{\lambda}\frac{\langle J_{\lambda}\{p_k=\sum_{i=1}^Nz_i^k\} \rangle}
{\langle J_{\lambda}, J_{\lambda}\rangle_{\beta}}J_{\lambda}\{p\},
\end{equation}
where
\begin{eqnarray}
\langle J_{\lambda}\{p_k=\sum_{i=1}^Nz_i^k\}\rangle=\prod_{i=1}^N\int_{0}^{+\infty} dz_i
\Delta(z)^{2\beta} J_{\lambda}\{p_k=\sum_{i=1}^Nz_i^k\}.
\end{eqnarray}

Comparing (\ref{expaninjack}) with (\ref{expaninjack1}), we have
\begin{eqnarray}\label{jackaverage}
\langle J_{\lambda}\{p_k=\sum_{i=1}^Nz_i^k\}\rangle=
\frac{J_{\lambda}\{p_k=N\}J_{\lambda}\{p_k=N+\beta^{-1}-1\}}
{J_{\lambda}\{p_k=\beta^{-1}\delta_{k,1}\}}.
\end{eqnarray}
The formula (\ref{jackaverage}) was conjectured in Ref. \cite{Cassia2020} and checked for the partitions
$\lambda$ with $|\lambda|\leq 9$. We have proved this conjecture here.

\section{Conclusions}
It is well known that the hermitian one-matrix model can be represented as an integrated CFT expectation value.
We have constructed the operators $\tilde W_{n}$ (\ref{Mna}) in terms of the generators of the Heisenberg algebra.
By inserting them into the integrated CFT expectation value, we have derived the constraints which contain
the $A$ and $B$-type Lassalle constraints. We explored the intrinsic connection between the constraint operators
and $W$-representations of the Gaussian hermitian one-matrix model (in the external field), $N\times N$ complex
and Hurwitz-Kontsevich matrix models via rescaling variable transformations. Similarly, through the constructed operators,
we have derived the desired constraints for the $\beta$-deformed hermitian matrix model, where two of the constraint operators
are associated with the $W$-representations of $\beta$-deformed Gaussian hermitian and $N\times N$ complex matrix models
via rescaling variable transformations. We investigated the superintegrability for the Gaussian hermitian one-matrix model
in the external field and $\beta$-deformed $N\times N$ complex matrix model. Based on the $W$-representations of these two
matrix models, we directly derived the corresponding character expansions with respect to the Schur functions and Jack
polynomials, respectively. Moreover the conjectured formula for the averages of Jack polynomials (\ref{jackaverage})
in Ref. \cite{Cassia2020} has been proved in this paper. For further research, it would be interesting to study whether
there exist $W$-representations of new matrix models associating with the rescaling constraint operators presented in this paper.
In addition, we have presented the generalized Virasoro constraints with higher algebraic structures
for the ($\beta$-deformed) hermitian one-matrix models. Their properties and applications still deserve further study.

In deriving the constraints for the $\beta$-deformed matrix model, we have obtained the second order total derivative operators
$\bar {\mathcal{W}}_{n}$ (\ref{barMn}) with respect to the integration variables. For the real Laughlin wave function
which is the ground state eigenfunction of the $A_{N-1}$-Calogero model, it is annihilated by the combination of $\bar {\mathcal{W}}_{n}$
and the generators of Virasoro algebra. It is also noted that the ground state eigenfunction of the many body problem (\ref{Le}),
i.e., power of the Vandermonde determinant, is annihilated by $\bar {\mathcal{W}}_{n}$.
Furthermore, we have constructed the operators $\mathscr{H}_n$ (\ref{GHS}) given by the combination of the Euler operator and
${\mathscr{\bar W}}_{n}$ which are the extended operators of $\bar {\mathcal{W}}_n$.
Some well known operators are included in $\mathscr{H}_n$ as special cases, such as the Laplace-Beltrami and Lassalle operators
(with the exchange operator). We have presented the eigenfunctions of the operators $\mathscr{H}_n$.
It is worthwhile to point out that exploring more applications of the total derivative operators $\bar {\mathcal{W}}_{n}$ and the
extended operators $\mathscr{H}_n$ would be interesting for further research.

\section *{Acknowledgments}

This work is supported by the National Natural Science Foundation of China (Nos. 11875194 and 12105104)
and the Fundamental Research Funds for the Central Universities, China (No. 2022XJLX01).


\end{document}